# Antiferromagnetic Correlations and the Pseudogap in HTS Cuprates

Jeffery L. Tallon

Industrial Research Ltd., P.O. Box 31310, Lower Hutt, New Zealand. E-mail: J.Tallon@irl.cri.nz

**Abstract:** Evidence is presented from $1/T_1T$ NMR and ARPES data for the sudden disappearance of 2D antiferromagnetic (AF) correlations in the lightly overdoped region ($p \approx 0.19$) at the T=0 metal-insulator transition where the pseudogap energy falls to zero. AF fluctuations thus appear to be intimately associated with the pseudogap and serve primarily to weaken superconductivity, strongly reducing the condensation energy and superfluid density.

## INTRODUCTION

The high-$T_c$ superconducting (HTS) cuprates exhibit a generic phase behaviour as a function of hole concentration, p, ranging from an antiferromagnetic (AF) insulator at zero doping to a metallic Fermi liquid at high doping with the appearance of superconductivity (SC) at intermediate doping levels. In spite of the disappearance of the 3D Neél state prior to the onset of SC 2D AF correlations are observed to persist well out into SC compositions [1]. In addition, underdoped cuprates exhibit normal state (NS) correlations above $T_c$ which result in a depletion of the density of states (DOS) referred to as the *pseudogap* [2]. While many models have been proposed for understanding the pseudogap a widely favoured scenario is that it arises from incoherent pairing fluctuations above $T_c$ [3], an outlook not shared by the present author. In view of the lack of consensus as to the nature of the pseudogap and the origin of the pairing interaction a number of key questions may be asked: (i) is the progression from a strongly AF-correlated underdoped phase to a Fermi liquid smooth or discontinuous? (ii) should the HTS cuprates in the SC region be regarded as doped antiferromagnets or AF Fermi liquids? (iii) is the disappearance of the pseudogap gradual (as might occur in a phase fluctuation model) or sudden (as for a competing correlation)? (iv) what is the relation, if any, between the pseudogap and AF correlations? The lack of such basic comprehension of the experimental situation goes a long way to understanding why so many mutually exclusive theoretical models of cuprate superconductivity are still extant [4]. We address these questions here.

## AF CORRELATIONS AND THE PSEUDOGAP

It is widely believed that $1/^{63}T_1T$ provides a clear measure of AF correlations, where $1/^{63}T_1$ is the copper spin-lattice relaxation rate given by

$$1/T_1T \sim \sum_{\mathbf{q}} |A_q|^2 \; \chi''(\mathbf{q},\omega_o) / \omega_o. \qquad (1)$$

$A_{\mathbf{q}}$ are the hyperfine coupling form factors, $\omega_o$ is the NMR frequency and $\chi''(\mathbf{q},\omega_o)$ is the imaginary part of the dynamic spin susceptibility. Typical data for underdoped cuprates show that $1/^{63}T_1T$ has a Curie-like 1/T dependence at high T associated with AF correlations but then falls at lower T due to the opening of the pseudogap [5]. The maximum occurs at T* which is widely viewed as the temperature at which the spin gap opens. Millis, Monien and Pines [6] introduced a phenomeno-logical expression for $\chi(\mathbf{q},\omega_o)$ which is enhanced at the AF wave vector $\mathbf{q} = \mathbf{Q}_{AF} \equiv (\pi,\pi)$. Inserting this in the expression for $1/T_1T$ and assuming that the AF correlation length $\xi^2 \gg 1$ one finds $1/^{63}T_1T \approx a_1\chi_s\,\tau_{SF}$ where $\chi_s$ is the static spin susceptibility and $\tau_{SF}$ is the AF spin fluctuation lifetime. The Curie-like T-dependence of $1/^{63}T_1T$ at high T thus implies that $\tau_{SF} \sim 1/T$ and that $1/^{63}T_1 \approx a_2\chi_s$. This appears to be precisely satisfied in the case of Y-124 [7]. At the same time one finds $1/^{17}T_1T \approx a_3\chi_s$ and finally the Knight shift, $K_s \approx a_3\,\chi_s + \sigma$ where $\sigma$ is the chemical shift. Thus

$$1/^{63}T_1 \sim 1/^{17}T_1T \sim (K_s - \sigma) \sim \chi_s. \tag{2}$$

Again, these relationships are well satisfied for Y-124 [7]. The characteristic T-dependence of $1/^{63}T_1T$ can thus be seen to derive from the 1/T dependence of $\tau_{SF}$ and the T-dependence of $\chi_s$ which, like $S^{el}/T$ (with $S^{el}$ = electronic entropy), is progressively and smoothly depressed with decreasing T due to the pseudogap. T* thus loses its meaning as a well-defined point at which a spin gap opens.

Experimentally, $1/^{63}T_1T$ maintains its high-T Curie-like T-dependence across the entire overdoped region in La-214 [5] thus suggesting that AF correlations also persist across the overdoped region. However, we argue that this inference is not justified. In La-214 (and less markedly in other cuprates) $\chi_s$ itself is found to develop an increasing 1/T dependence in the overdoped region [8] possibly due to the proximity of the van Hove singularity. By reference to equ. (2) this could account for the persisting 1/T dependence of $1/^{63}T_1T$, i.e this dependence in the overdoped region could derive from $\chi_s$ rather than $\tau_{SF}$. A more robust measure of AF correlations is the ratio $^{17}T_1/^{63}T_1$ in which the effect of $\chi_s$ is divided out. From the above considerations $^{17}T_1/^{63}T_1 \sim \tau_{SF}$ but more generally,

$$^{17}T_1 / {}^{63}T_1 \sim \langle 1 + f_{\mathbf{q}}^2 \rangle \approx 1 + a_3\,\xi^2 = 1 + C_{AF}\,T^{-1} \tag{3}$$

where the average is over $\mathbf{q}$. Here $f_{\mathbf{q}}$ is the ratio of the enhanced AF susceptibility to the bare FL susceptibility and $C_{AF}$ is a measure of the AF correlations. The ratio $^{17}T_1/^{63}T_1$ has been determined [9] by Takigawa *et al.* for Y-123 at two doping levels and by Tomeno *et al.* for Y-124 and the T-dependence of eqn. (3) is found. We have fitted the data to obtain the p-dependent parameter $C_{AF}(p)$ and this is plotted in Fig. 1. This is seen to fall sharply to zero at the critical doping point of p=0.19 precisely where the pseudogap energy, $E_g$, determined from NMR and heat capacity falls to zero, as shown by the inset. Values of p are determined from $\delta$ values in $YBa_2Cu_3O_{7-\delta}$ or from the roughly parabolic variation of $T_c$ with p which may be approximated by $T_c = T_{c,max} \times [1 - 82.6(p-0.16)^2]$ [10].

One major effect of AF correlations is to heavily reduce quasiparticle (QP) lifetimes near the zone boundary at k=(π,0) due to scattering from spin fluctuations. This may be seen in the suppression of the NS QP peaks in ARPES spectra at the FS crossing near (π,0) but not at the FS crossing on the zone diagonal near (0,0) [11]. If one focuses on the spectra near (π,0) at about 100K, i.e. above $T_c$, then underdoped samples show the suppression of the QP peak as well as the pushing back of the leading edge due to the NS pseudogap. In contrast, overdoped samples exhibit a closure of the pseudogap (the mid-point of the leading edge coincides with the Fermi energy) and the recovery of the QP peak. We have examined the ARPES spectra at 100K of 11 Bi-2212 samples with different doping states from the Stanford and Chicago groups and summarise the data in Fig. 2. This, again, shows the pseudogap energy falling to zero [12] near p=0.19 and the abrupt recovery of the QP peak at the same point. The $T_c$ values are plotted as open squares for all spectra with suppressed QP peaks and as solid squares where the QP is fully recovered. There is a sudden recovery at p=0.19 as indicated by the spectra shown in the figure either side of this point [11,12]. Also shown in Fig. 1 (open circles) are the values of $T_{min}$ where the resistivity of $La_{2-x}Sr_xCuO_4$ crosses over from metallic to semiconducting. The point where $T_{min} \to 0$ is the metal/insulator transition and it clearly coincides with the disappearance of both the pseudogap and AF correlations.

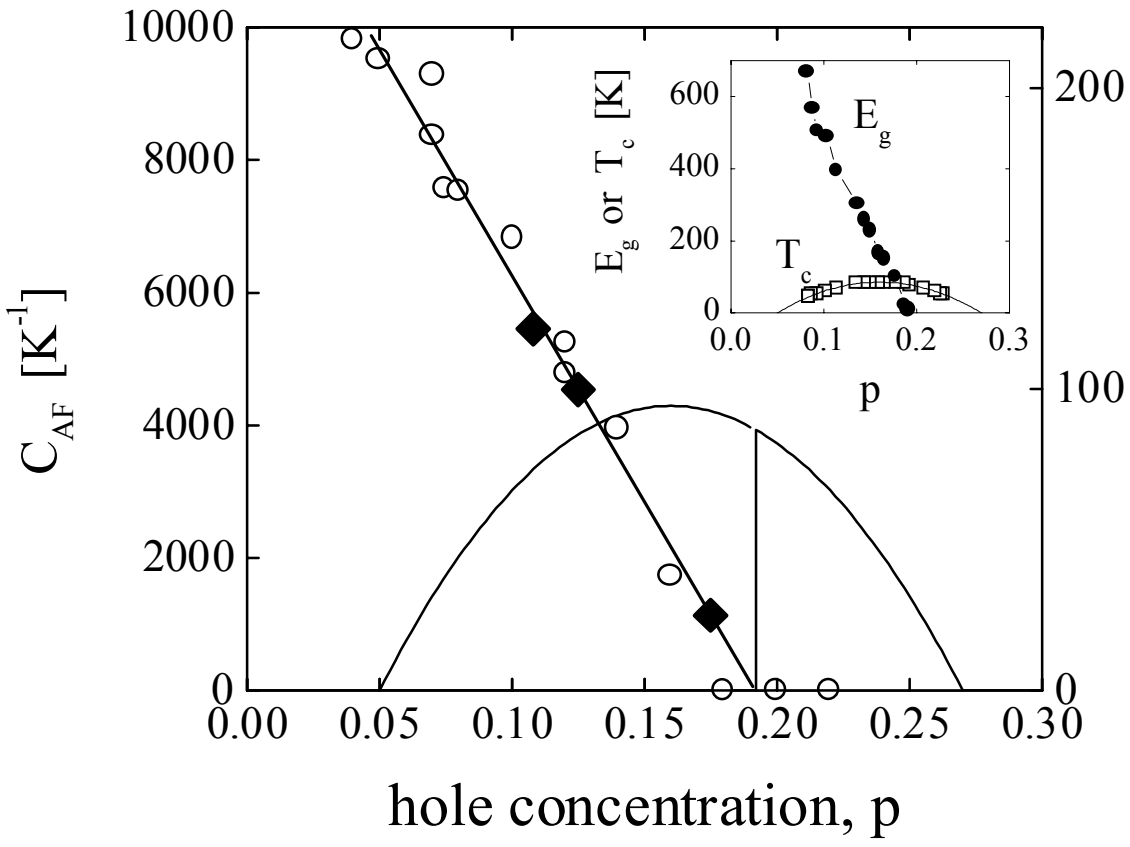

Fig. 1. The p-dependence of the AF parameter $C_{AF}$ (◆) and crossover temperature $T_{min}$ (O). Inset: the p-dependence of $T_c$ and the pseudogap energy $E_g$ for $Y_{0.8}Ca_{0.2}Ba_2Cu_3O_{7-\delta}$.

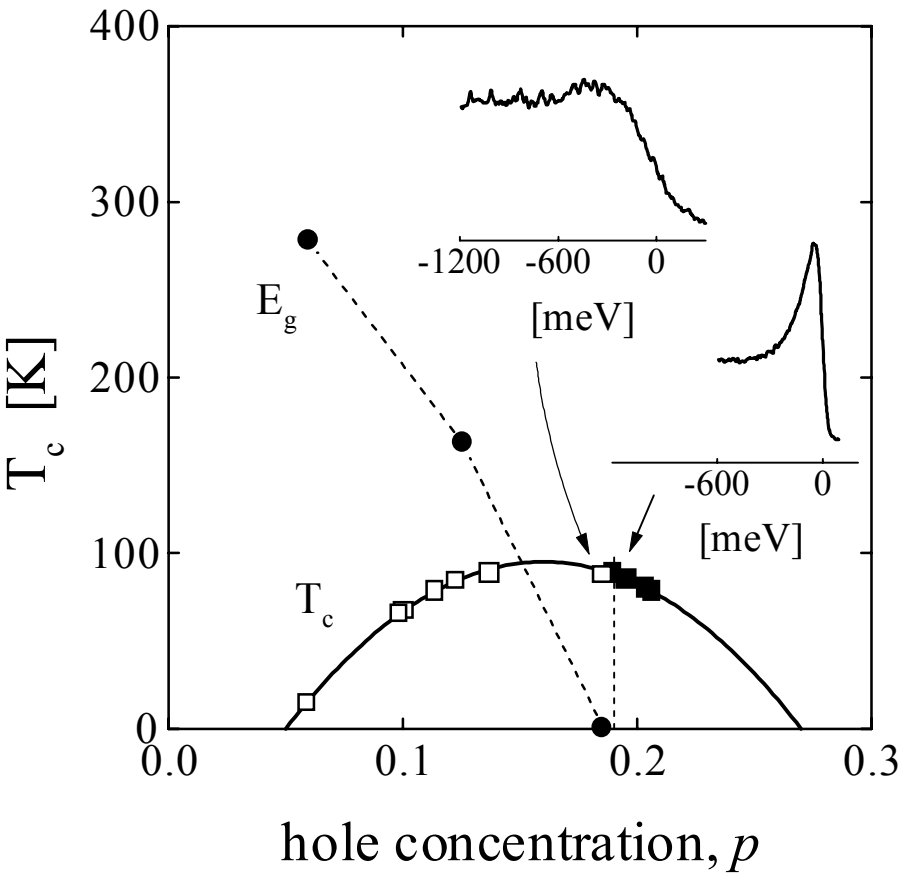

Fig. 2. The p-dependence of $E_g$ and $T_c$ from ARPES. Open squares: no quasi-particle (QP) peak. Filled squares: full QP peak as shown in the two insets.

The above results reveal the *sudden* disappearance of AF correlations at the critical doping state p=0.19, just where the pseudogap disappears and at the location of the metal-insulator transition at T=0. At this point the T=0 condensation energy, superfluid density and critical currents all pass through a sharp maximum and this has been interpreted within a quantum critical point scenario [13].The sudden loss of AF correlation is further borne out by inelastic neutron scattering which, for fully oxygenated Y-123, shows only a weak enhancement in susceptibility at $\mathbf{q}=\mathbf{Q}_{AF}$ that barely rises above background [1]. Moreover, Rübhausen *et al* [14] have observed a sudden loss of the 2000 $cm^{-1}$ two-magnon Raman scattering peak in the lightly overdoped region (at $p\approx0.20$). This peak, observed in underdoped and optimally-doped samples, is attributed to a photon-induced two-magnon excitation in an AF background. Its demise indicates the destruction of the AF background. These results indicate that AF correlations are intimately associated with the pseudogap which however is more than just spin correlations. It is known from heat capacity that both spin and charge degrees of freedom freeze out equally with the establishment of the pseudogap state [2]. Spin- and charge-ordered stripes are then a possible scenario. Antiferromagnetism would appear only to weaken and suppress SC with progressive underdoping and the question has to asked whether spin fluctuations can be responsible for pairing if they are substantially suppressed over much of the overdoped side.